\documentclass{elsarticle}
\usepackage{amssymb}

\usepackage[figuresright]{rotating}

\begin{document}

\begin{frontmatter}




\title{ Localization or tunneling in asymmetric double-well potentials}


\author{Dae-Yup Song}

\address{Department of Physics Education, Sunchon National University, Jeonnam 57922, Korea}

\begin{abstract}
An asymmetric double-well potential is considered, assuming that the wells are parabolic around the minima. The WKB wave function of a given energy is constructed inside the barrier between the wells.  By matching the WKB function to the exact wave functions of the parabolic wells on both sides of the barrier,  for two almost degenerate states, we find a  quantization condition for the energy levels which reproduces the known energy splitting formula between the two states.
For the other low-lying non-degenerate states, we show that the eigenfunction should be primarily localized in one  of the wells with negligible magnitude in the other. Using Dekker's method [Physica 146A (1987) 375], the present analysis generalizes earlier results for weakly biased double-well potentials to systems with arbitrary asymmetry.
\end{abstract}

\begin{keyword}
Quantum tunneling \sep  Localization  \sep  Double-well potential

\end{keyword}

\end{frontmatter}



\section{ Introduction}
\label{Introduction}


Quantum tunneling has been of continuing interest since the advent of quantum mechanics, and the inversion of an ammonia molecule and proton tunneling are  well-known examples of  microscopic quantum tunneling which may be described by one-dimensional models (see, e.g., Refs.~\cite{BCM,Gordon,NKSSK}). For an one-dimensional symmetric double-well potential of two wells being sufficiently separated and deep, the lower energy eigenvalues are closely bunched in pairs, to give rise to tunneling dynamics for a wave packet initially localized in one of the wells with the energy of approximately one of the eigenvalues  (see, e.g., Ref.~\cite{DekkerPRA}). It has  then been well-known that, upon adding a small asymmetry to a symmetric potential, the two states that started out as tunneling states in the symmetric case correspond increasingly to states localized in one well or the other, to quench the tunneling motion \cite{BCM,Gordon,NKSSK}. Further, it is known that, for an asymmetric potential, if there exist two states which are almost degenerate then tunneling dynamics take place for a localized wave packet made up of those states \cite{WT}, and the analytic expression for the energy splitting between the states is given in Ref.~\cite{JPSADLW}.

In addition to microscopic quantum tunneling,  a recent breakthrough makes it possible to realize  macroscopic quantum tunneling  in a superconducting quantum interference device with Josephson junctions where  numerous microscopic degrees of freedom are  tied together to form a collective dynamical variable (see, e.g., Refs.~\cite{JPSADLW,Koch,CSDG}), and, for the system of the  transmon Hamiltonian of a cosine potential,  the analytic expression for the energy splitting between the nearly degenerate even and odd eigenstates is given  when the amplitude of the potential is large \cite{CSDG}.
In obtaining the expression for the splitting \cite{CSDG} that could exactly reproduce the rigorous mathematical expression for the widths of the low-lying energy bands of the associated Mathieu equation (see Refs.~\cite{Blanch,CUMS,nist,Koch}),  a WKB wave function is normalized to be matched in a forbidden region of one of the wells onto a normalized eigenfunction of an harmonic oscillator centered at the minimum of the well; this normalization  corrects the underestimate  for a low-lying state that the method described in Ref.~\cite{LL} may give.
Then, the tight bonding approximation for a periodic system is applied based on the normalized WKB wave functions \cite{CSDG}. 
For  an asymmetric double-well potential, while the analytic expression for the energy splitting between pairwise degenerate left and right states is succinctly given in Ref.~\cite{JPSADLW}, before applying the  Lifshitz-Herring approximation for the expression, it is necessary to normalize the WKB wave function by matching it onto a normalized eigenfunction of a  harmonic oscillator in a forbidden region between the wells  \cite{JPSADLW,DekkerPRA,Song}.

Instead of using an approximation with the normalized WKB functions, by assuming that the two wells are parabolic with an angular  frequency $\omega_0$, Dekker  in Ref.~\cite{DekkerA} first shows  that  the consistency condition which comes from that a WKB function should match onto the exact solutions on both sides of the potential barrier determines the energy splitting between the pair of the lowest eigenvalues, when a small bias is added to a symmetric double-well potential.  Further, he shows, if the bias increases, the ground state becomes almost localized in the deeper of the two wells, implying that the tunneling motion is quenched. In Ref.~\cite{Song}, a similar analysis is carried out for the pairs of the low-lying states of an asymmetric potential assuming that  the difference of the minima is close to a multiple of $\hbar\omega_0$. Indeed, in the region where the potential is quadratic(parabolic), the exact wave function is described by the parabolic cylinder function, and thus the wave function  in its asymptotic expansion could have an additional leading term   aside from that of a Hermite-Gaussian wave function of a harmonic oscillator \cite{WW}, which makes Dekker's method work.

In this article, we will consider an asymmetric double-well potential for which the separation between the two wells is large,  assuming that the two wells are parabolic with angular frequencies $\omega_L$ and $\omega_R$ around the minima of the left-hand and right-hand wells, respectively. In the case that  there exist two almost degenerate eigenstates, through the systematic application of Dekker's method, we arrive at a quantization condition for the eigenvalues which is a refined version of that found in a semiclassical analysis without the  assumption of parabolicity  \cite{FF,Connor}. We further analyze this quantization condition, to find the energy splitting formula given in Ref.~\cite{JPSADLW}.
For a low-lying non-degenerate eigenstate, however, as a leading term in the asymptotic expansion of the  wave function is dominant over  that of the Hermite-Gaussian wave function in one of the wells, we only have an equation which shows that the eigenstate has negligible magnitude in the well and  is primarily localized in the other. We also include the two-level approach  which amounts to the method used in Ref.~\cite{JPSADLW}, not only to provide the explicit form of the normalized WKB function (see, e.g., Refs.~\cite{Song,Rastelli}) but also to supplement the validity of  Dekker's method in the vanishing limit of  the energy splitting where  the wave function is accurately described by the Hermite-Gaussian functions.

This paper is organized as follows: In Section \ref{sec:wave}, using  Dekker's method, we find the quantization condition and  the energy splitting formula for the  two almost degenerate eigenstates. For a low-lying non-degenerate eigenstate, an equation which shows the localization of the state is given. Particular attention is paid on the validity of the method.  In Section \ref{sec:two}, the two-level approximation is used to re-obtain the  energy splitting formula. In Section \ref{sec:concluding}, we give some concluding remarks.

\section{ A WKB wave function and the quantization condition }
\label{sec:wave}

\begin{figure}
\centering
\includegraphics[width=8cm]{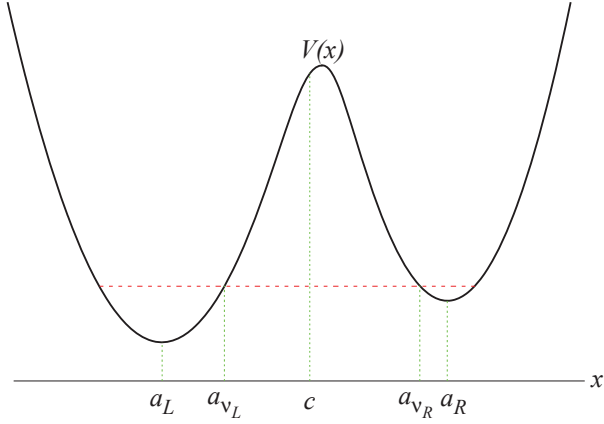}
\caption{  An asymmetric double-well potential $V(x)$ and the classical turning points $a_{\nu_L}$ and $a_{\nu_R}$. We assume that $V(x)$ is quadratic around its minima at $x=a_L$ and at $x=a_R$ with angular frequencies $\omega_L$ and $\omega_R$, respectively. }
\label{figure1}
\end{figure}

We assume that the double-well potential $V(x)$ has quadratic mimima at $x=a_L$ and at $x=a_R$ with angular frequencies $\omega_L$  and   $\omega_R$, respectively (see Fig.~\ref{figure1}). For the eigenfunction $\psi(x)$ corresponding to the eigenvalue 
\begin{equation}
E=V(a_L )+ (\nu_L+ \frac{1}{2})\hbar\omega_L=V(a_R )+ (\nu_R+ \frac{1}{2})\hbar\omega_R,
\label{defenergy}
\end{equation}
the Schr\"{o}dinger equation
\begin{equation}
H\psi(x)=-\frac{\hbar^2}{2m} \frac{d^2}{dx^2}\psi(x) +V(x) \psi(x)= E\psi(x)
\label{Schroedinger}
\end{equation}
is thus written in the quadratic region around $a_i$ as
\begin{equation}
-\frac{\hbar^2}{2m} \frac{d^2}{dx^2}\psi(x) +\frac{m\omega_i^2}{2}(x-a_i)^2\psi(x)= \hbar\omega_i (\nu_i + \frac{1}{2})\psi(x),
\label{sew}
\end{equation}
where $i$ denotes $L$ or $R$, with the particle's mass $m$.
We also assume that the potential is monotonically decreasing for $x<a_L$ and monotonically increasing for $x>a_R$, so that energy spectrum is discrete with square-integrable eigenfunctions.

As we are interested in the barrier penetration, we restrict our attention on non-negative  $\nu_L$  and $\nu_R$.
By introducing 
\begin{equation}
 z_i=\frac{\sqrt{2}(x-a_i)}{l_i},
\end{equation}
with
\begin{equation}
l_i=\sqrt{{\hbar}\over{m\omega_i}},
\end{equation}
we rewrite Eq.~($\ref{sew}$)  as
\begin{equation}
\label{pcfrl}
\frac{d^2 \psi}{d z_i^2}+ \left(\nu_i+\frac{1}{2} -\frac{z_i^2}{4}\right)\psi =0.
\end{equation}
The solutions of Eq. (\ref{pcfrl}) are parabolic cylinder  functions and we write the wave function $\psi(x)$ as
\begin{equation}
\psi_L(x)=C_L D_{\nu_L}(- z_L)=C_L D_{\nu_L}\left(\frac{\sqrt{2}(a_L -x)}{l_L}  \right)
\label{pcfl}
\end{equation}
and
\begin{equation}
\psi_R(x)=C_R D_{\nu_R}( z_R)=C_R D_{\nu_R}\left(\frac{\sqrt{2}(x-a_R)}{l_R}  \right)
\label{pcfr}
\end{equation}
around the minima of the left-hand well and right-hand well, respectively, with constant $C_L$ and $C_R$ .  
The choice of the solution in Eq.~($\ref{pcfl})$ (in Eq.~($\ref{pcfr}$)) is made bearing in mind that we wish to construct a normalizable wave function so that $\int_{-\infty}^{a_L} |\psi_L(x)|^2 dx$ ($\int_{a_R}^\infty |\psi_R(x)|^2 dx$)  is finite  if we suppose the expression of $\psi_L(x)$ ($\psi_R(x)$) is valid for $x< a_L$ ($x>a_R$).

$D_\alpha (z)$ of Whittaker can be expressed in terms of other parabolic cylinder functions $U$ and $V$ as \cite{AS}
\begin{equation}
D_\alpha (z)= \cos(\alpha \pi)U(-\alpha -\frac{1}{2}, -z) + \frac{\pi}{\Gamma(-\alpha)}V(-\alpha -\frac{1}{2}, -z).
\label{pcfrel}
\end{equation}
Equation (\ref{pcfrel}) can be used to find the asymptotic expansion for $|z|\gg|\alpha|$ 
\begin{eqnarray}
D_\alpha (z)~ & \sim&~ \cos(\alpha\pi)\exp\left(-\frac{z^2}{4}\right)|z|^\alpha\left[1-\frac{\alpha(\alpha-1)}{2z^2}+\cdots\right]
\nonumber\\
&&~+ \frac{\sqrt{2\pi}}{\Gamma(-\alpha)|z|^{\alpha+1}}\exp\left(\frac{z^2}{4}\right)
\left[1+\frac{(\alpha+1)(\alpha+2)}{2z^2}+\cdots\right]
\label{AsymD}
\end{eqnarray}
which is completely valid {\em only for $z$ real and negative} \cite{AS}. 
Using the relation
\[\Gamma(-\alpha)=-\frac{\pi}{\alpha!\sin( \alpha\pi)},\] 
Eq.~($\ref{AsymD}$) is then rewritten as (see also Ref.~\cite{MG})
\begin{eqnarray}
D_\alpha (z)~ & \sim&~ \cos(\alpha\pi)|z|^\alpha\exp\left(-\frac{z^2}{4}\right)\left[1-\frac{\alpha(\alpha-1)}{2z^2}+\cdots\right]
\nonumber\\
&&~-\frac{\sin(\alpha\pi)\alpha !}{|z|^{\alpha+1}} \sqrt{ \frac{2}{\pi}}\exp\left(\frac{z^2}{4}\right)
\left[1+\frac{(\alpha+1)(\alpha+2)}{2z^2}+\cdots\right].
\label{AsymDf}
\end{eqnarray}

In the quadratic region of the left-hand well satisfying $\sqrt{2}(x-a_L)\gg \nu_L l_L$, using Eqs.~(\ref{pcfl}) and (\ref{AsymDf}), we obtain
\begin{eqnarray}
&\psi_L(x)
\simeq& C_L\cos(\nu_L \pi)\left(\frac{\sqrt{2}(x-a_L)}{l_L}\right)^{\nu_L}e^{-\frac{(x-a_L)^2}{2l_L^2}}\nonumber\\
&&-C_L\sin(\nu_L \pi)\nu_L !\sqrt{\frac{2}{\pi}}\left(\frac{l_L}{\sqrt{2}(x-a_L)}\right)^{\nu_L+1}e^{\frac{(x-a_L)^2}{2l_L^2}}.
\label{DL}
\end{eqnarray}
In the region of the right-hand well of $\sqrt{2}(a_R-x) \gg\nu_R l_R$, using Eqs.~(\ref{pcfr}) and (\ref{AsymDf}),  we also have
\begin{eqnarray}
&\psi_R(x)
\simeq& C_R\cos(\nu_R \pi)\left({\sqrt{2}(a_R-x)\over l_R}\right)^{\nu_R}e^{-\frac{(x-a_R)^2}{2l_R^2}}\nonumber\\
&&-C_R\sin(\nu_R \pi)\nu_R !\sqrt{\frac{2}{\pi}}\left(\frac{l_R}{\sqrt{2}(a_R-x)}\right)^{\nu_R+1}e^{\frac{(x-a_R)^2}{2l_R^2}}.
\label{DR}
\end{eqnarray}
To be precise, in obtaining  Eqs.~(\ref{DL}) and (\ref{DR}), we have assumed the existence of certain numbers $N_L~( \gg \nu_L)$ and $N_R~( \gg \nu_R)$ such that the potential is quadratic in the regions  
$a_L \leq x \leq a_L+N_Ll_L$ and $a_R-N_Rl_R \leq x  \leq a_R$.
We note that, in Eq.~(\ref{DL}), the magnitude of the ratio of the first term to the second one  at $x=a_L +N_Ll_L$ is 
$\sqrt{\frac{\pi}{2}}\frac{1}{\nu_L!}|\frac{\cos(\nu_L \pi)}{\sin (\nu_L\pi)}|(\sqrt2 N_L)^{2\nu_L+1}e^{-N_L^2},$
so that, if $\frac{1}{|\sin(\nu_L \pi)|}$ is moderate(not exponentially large), the  magnitude of the first term for $N_L$ large is  exponentially small compared to that of the second term  when $x\gtrsim a_L +N_Ll_L$. Likewise, in Eq.~(\ref{DR}),  if  $N_R$ is  large and   $\frac{1}{|\sin(\nu_R\pi)|}$  is moderate, the second term is dominant over the first one for $x\lesssim a_R-N_Rl_R$.

In the classically forbidden region between the wells, with a point $c$ satisfying $a_L<c<a_R$, the WKB approximation to an eigenfunction is written as
\begin{equation}
\psi_{WKB}(x)=A\sqrt{\frac{\hbar}{p(x)}}\exp\left(\int_c^x \frac{p(y)}{\hbar}dy\right)
            + B\sqrt{\frac{\hbar}{p(x)}}\exp\left(-\int_c^x \frac{p(y)}{\hbar}dy\right),
\end{equation}
where $p(x)$ is defined as
\begin{equation}
p(x)= \sqrt{2m[V(x) -E]}.
\end{equation}
We here define the classical turning points $a_{\nu_L}$ and $a_{\nu_R}$ satisfying $a_L<a_{\nu_L}<c<a_{\nu_R}<a_R$   (see Fig.~\ref{figure1}) and
\begin{equation}
E=V(a_{\nu_L})=V(a_{\nu_R}).
\end{equation}
In the region of quadratic potential near the left minimum, by introducing 
\begin{eqnarray}
W_L(\nu_L) &=& \int_{a_{\nu_L}}^x \frac{p(y)}{\hbar}dy =\int_{a_{\nu_L}}^x 
\sqrt{ \frac{(a_L -y)^2}{l_L^2}- 2\nu_L-1}~dy \cr
&=&\int_{\sqrt{2\nu_L+1}}^{\frac{-a_L+x}{l_L}}\sqrt{z^2-2\nu_L-1}~dz
\end{eqnarray}
 for $x>a_{\nu_L}$, we have the asymptotic expansion 
\begin{eqnarray}
W_L(\nu_L)&=&\frac{(-a_L+x)^2}{2l_L^2} -\frac{1}{2}(\nu_L+\frac{1}{2}) \cr
&&+(\nu_L+\frac{1}{2})\ln\frac{\sqrt{2}(-a_L+x)}{l_L\sqrt{\nu_L+\frac{1}{2}}}
+O\left(\frac{l_L^2(\nu_L+\frac{1}{2})}{(-a_L+x)^2}\right).
\label{W_L}
\end{eqnarray}
With 
\[\int_{c}^x \frac{p(y)}{\hbar}dy=\int_c^{a_{\nu_L}} \frac{p(y)}{\hbar}dy+W_L(\nu_L),\]
 we then find
\begin{eqnarray}
&&\psi_{WKB}(x)\cr
&&\simeq A\frac{\sqrt{l_L}}{\pi^{\frac{1}{4}}}\sqrt{\nu_L!g_{\nu_L}}
\left(\frac{l_L}{\sqrt{2}(x-a_L)}\right)^{\nu_L+1}
\exp\left(\frac{(-a_L+x)^2}{2l_L^2}- \int_{a_{\nu_L}}^c \frac{p(y)}{\hbar}dy\right) \phantom{\hspace{0.4cm}} 
\cr
&&~~+B\frac{(4\pi)^{\frac{1}{4}}\sqrt{l_L}}{\sqrt{\nu_L!g_{\nu_L}}}
\left(\frac{\sqrt{2}(x-a_L)}{l_L}\right)^{\nu_L}
\exp\left(-\frac{(-a_L+x)^2}{2l_L^2}+\int_{a_{\nu_L}}^c \frac{p(y)}{\hbar}dy\right)~
\end{eqnarray}
 for $\sqrt{2}(-a_L+x)\gg l_L\sqrt{2\nu_L+1}$, where 
\begin{equation}
g_k =\frac{\sqrt{2\pi}}{k!}\left(k+\frac{1}{2}\right)^{k+\frac{1}{2}}e^{-k-\frac{1}{2}}.
\end{equation}
Matching this expression of $\psi_{WKB}(x)$ onto the asymptotic form of $\psi_L(x)$ in the overlap region, we have
\begin{eqnarray}
A&=&- \sin(\nu_L \pi)\frac{\sqrt{2\nu_L!}}{{\pi^{\frac{1}{4}}\sqrt{l_L g_{\nu_L}}}}
\exp\left(\int_{a_{\nu_L}}^c \frac{p(y)}{\hbar}dy\right) C_L,
\label{ACL}
\\
B&=&\cos(\nu_L\pi)\frac{\sqrt{\nu_L!g_{\nu_L}}}{(4\pi)^{\frac{1}{4}}\sqrt{l_L}}
\exp\left(-\int_{a_{\nu_L}}^c \frac{p(y)}{\hbar}dy\right)C_L.
\label{BCL}
\end{eqnarray}

In passing, we note that the form of $g_k$ first appears in Ref.~\cite{Furry} from the comparison between an exact eigenfunction and  the corresponding 
 semiclassical wave function  for the harmonic oscillator. For a plane pendulum in quantum mechanics described by Mathieu's equation, it is well-known that, in the limit of low probability for barrier 
penetration, the correct asymptotic mathematical expression  for the width 
of the $k$-th energy band ($k$-th stable region) is given by the product of   $g_{k-1}$ and the result obtained from the  semiclassical calculation  \cite{CUMS} at the leading order \cite{nist}.

With the notation
\begin{eqnarray}
W_R(\nu_R)&=&\int_x^{a_{\nu_R}} \frac{p(y)}{\hbar}dy=\int_x^{a_{\nu_R}} 
\sqrt{ \frac{(a_R -y)^2}{l_R^2}- 2\nu_R-1}~dy  \cr
&=&\int_{\sqrt{2\nu_R+1}}^{\frac{a_R-x}{l_R}}\sqrt{z^2-2\nu_R-1}~dz,
\end{eqnarray}
using the asymptotic relation
\begin{equation}
W_R(\nu_R)\simeq\frac{(a_R-x)^2}{2l_R^2}-\frac{1}{2}(\nu_R+\frac{1}{2})+(\nu_R+\frac{1}{2})\ln\frac{\sqrt{2}(a_R-x)}{l_R\sqrt{\nu_R+\frac{1}{2}}},
\end{equation}
in the quadratic-potential region  around the right minimum satisfying $\sqrt{2}(a_R-x)\gg\l_R\sqrt{2\nu_R+1}$,
we have
\begin{eqnarray}
&&\psi_{WKB}(x)  \cr
&&\simeq A\frac{(4\pi)^{\frac{1}{4}}\sqrt{l_R}}{\sqrt{\nu_R!g_{\nu_R}}}
\left(\frac{\sqrt{2}(a_R-x)}{l_R}\right)^{\nu_R}
\exp\left(-\frac{(a_R-x)^2}{2l_R^2}+\int_c^{a_{\nu_R}} \frac{p(y)}{\hbar}dy\right) \phantom{\hspace{1.2cm}}
\cr
&&~+B\frac{\sqrt{l_R\nu_R!g_{\nu_R}}}{\pi^{\frac{1}{4}}}
\left(\frac{l_R}{\sqrt{2}(a_R-x)}\right)^{\nu_R+1}
\exp\left(\frac{(a_R-x)^2}{2l_R^2}- \int_c^{a_{\nu_R}} \frac{p(y)}{\hbar}dy\right).
\end{eqnarray}
Comparing $\psi_{WKB}(x)$ and $\psi_R(x)$ in the overlap region, we obtain
\begin{eqnarray}
A&=& \cos(\nu_R\pi)\frac{\sqrt{\nu_R!g_{\nu_R}}}{(4\pi)^{\frac{1}{4}}\sqrt{l_R}}
\exp\left(-\int_c^{a_{\nu_R}} \frac{p(y)}{\hbar}dy\right)C_R,
\label{ACR}
\\
B&=&-\sin(\nu_R\pi)\frac{\sqrt{2\nu_R!}}{\pi^{\frac{1}{4}}\sqrt{l_Rg_{\nu_R}}}
\exp\left(\int_c^{a_{\nu_R}} \frac{p(y)}{\hbar}dy\right) C_R.
\label{BCR}
\end{eqnarray}

For the reasons given after Eq.~($\ref{DR}$), for large $N_L$ and moderate $\frac{1}{|\sin(\nu_L\pi)|}$, the first term in  Eq.~($\ref{DL}$)  is subleading in the overlap region, which means that Eq.~($\ref{BCL}$) comes from the comparison between  exponentially small terms compared to the leading ones. For the WKB approximation, it has been discussed that the accuracy of the approximation is not such as to allow the retention in the wave function of exponentially small terms superimposed on exponentially large ones \cite{LL}. This  implies that, while Eq.~($\ref{ACL}$) is valid within the approximation,  this approximation does not provide the validity of  Eq.~($\ref{BCL}$)  for large $N_L$ and moderate $\frac{1}{|\sin(\nu_L\pi)|}$. On the other hand, in the limit of  $\nu_L \rightarrow n_L$ with a non-negative integer $n_L$,  as the fact 
\begin{equation}
D_{n_L}(z)= 2^{-\frac{n_L}{2}}e^{-\frac{z^2}{4}}H_{n_L}\left( \frac{z}{\sqrt2}\right)
\end{equation} 
implies, $\psi_L(x)$ reduces to $(-)^{n_L}C_L\pi^{\frac{1}{4}}\sqrt{n_L!l_L}\psi_{n_L}^{sho}(a_L;x)$,
with  the $n_L$th excited state harmonic oscillator wave function
\begin{eqnarray}
&&\psi_{n_L}^{sho}(a_L;x)=\frac{1}{\pi^\frac{1}{4}\sqrt{2^{n_L}n_L!l_L}}
H_{n_L}\left(\frac{x-a_L}{l_L}\right)e^{-\frac{(x-a_L)^2}{2l_L^2}}\nonumber\\
&&~=\frac{1}{\pi^\frac{1}{4}\sqrt{2^{n_L}n_L!l_L}}e^{-\frac{(x-a_L)^2}{2l_L^2}}\left(\frac{2(x-a_L)}{l_L}\right)^{n_L}[1-\frac{n_L(n_L-1)l_L^2}{4(x-a_L)^2}+\cdots], \phantom{\hspace{0.7cm}} 
\label{sho}
\end{eqnarray}
where  $H_{n_L}$ denotes the $n_L$th order  Hermite polynomial.
Thus, if $\nu_L $ is in the (Hermite-Gaussian) limit of $\frac{1}{|\sin(\nu_L\pi)|}\rightarrow \infty$,
Eq.~($\ref{BCL}$) is valid while the WKB method may not provide the validity of Eq.~($\ref{ACL}$).
Likewise,  for large $N_R$, Eq.~(\ref{BCR}) is valid if $\frac{1}{|\sin(\nu_R\pi)|}$ is moderate, and Eq.~(\ref{ACR}) is valid if $\nu_R $ is in the  limit of $\frac{1}{|\sin(\nu_R\pi)|}\rightarrow \infty$, in the approximations.

Here,  we assume that the potential is quadratic in the region $|x-a_L|\le N_Ll_L$ with the large $N_L$; for small $n_L$, 
\begin{equation}
\epsilon_L=V(a_L) +(n_L+\frac{1}{2})\hbar\omega_L
\end{equation}
is thus approximately an eigenvalue of the double-well system so that, for the corresponding eigenvalue $E$,  $\nu_L\simeq n_L$. For this eigenvalue,  \[(-)^{n_L}C_L\pi^{\frac{1}{4}}\sqrt{n_L!l_L}\psi_{n_L}^{sho}(a_L;x)\]
is a good approximation to  $\psi_L(x)$, which implies that Eq.~($\ref{BCL}$) is valid.
Similarly, for a small non-negative integer $n_R$, 
\begin{equation}
\epsilon_R=V(a_R) +(n_R+\frac{1}{2})\hbar\omega_R
\end{equation}
is approximately an eigenvalue of the double-well system, and for the corresponding eigenvalue  Eq.~($\ref{ACR}$) is  valid with $\nu_R\simeq n_R$.

For $\nu_L\simeq n_L$  with moderate $\frac{1}{|\sin(\nu_R\pi)|}$,  using Eqs.~(\ref{BCL}) and (\ref{BCR}) which are (approximtely) valid in this case, we find  
\begin{equation}
\frac{C_L}{C_R}=-\frac{2\sin(\pi\nu_R)}{\cos(\pi\nu_L)}\sqrt{\frac{\nu_R!l_L}{\nu_L!l_R}}
\frac{1}{\sqrt{g_{\nu_L}g_{\nu_R}}}~e^{\int_{a_{\nu_L}}^{a_{\nu_R}} \frac{p(y)}{\hbar}dy}.
\label{LR2}
\end{equation}
For  $\nu_R\simeq n_R$ with moderate $\frac{1}{|\sin(\nu_L\pi)|}$, using Eqs.~($\ref{ACL}$) and ($\ref{ACR})$, likewise, we have 
\begin{equation}
\frac{C_L}{C_R}=-\frac{\cos(\pi\nu_R)}{2\sin(\pi\nu_L)}\sqrt{\frac{\nu_R!l_L}{\nu_L!l_R}}
\sqrt{g_{\nu_L}g_{\nu_R}}~e^{-\int_{a_{\nu_L}}^{a_{\nu_R}} \frac{p(y)}{\hbar}dy}.
\label{LR1}
\end{equation}

If  $\nu_L\simeq n_L$ with  moderate $\frac{1}{|\sin(\nu_R\pi)|}$,  while the validity of Eq.~($\ref{LR1}$) as well as that of Eq.~($\ref{ACR})$ is {\em not} provided by the the WKB approximation, Eq.~($\ref{LR2}$) is valid within the approximations   
to show that $|C_L|\gg |C_R|$. The ratio $R(\nu_L,\nu_R)$ of the probability of the particle being in the classically allowed region of the left-hand well to that of the right-hand well is given as
\[   R(\nu_L,\nu_R)= \bigg|\frac{C_L}{C_R}\bigg|^2 \frac{l_L}{l_R}
\frac{\int_{-\sqrt{4\nu_L+2}}^{\sqrt{4\nu_L+2}}[D_{\nu_L}(z_L)]^2 dz_L}{\int_{-\sqrt{4\nu_R+2}}^{\sqrt{4\nu_R+2}}[D_{\nu_R}(z_R)]^2 dz_R}.\]
Since $D_{\alpha}(z)$ is an entire function of $z$ with $D_\alpha(0)= \frac{2^{\frac{\alpha}{2}}}{\sqrt\pi} (\frac{\alpha-1}{2})! \cos(\frac{\alpha}{2}\pi)$ and $\frac{dD_\alpha(z)}{dz}|_{z=0}=\frac{2^\frac{\alpha+1}{2}}{\sqrt\pi}\left(\frac{\alpha}{2}\right)! \sin(\frac{\alpha}{2}\pi)$ \cite{WW}, for the large $N_L,$ $N_R$ and moderate $\frac{1}{|\sin(\nu_R\pi)|},$   Eq.~($\ref{LR2}$) thus implies that the eigenstate is primarily localized in the left-hand well with negligible magnitude for $x>c$.
We note that, in the expression of $\psi_R(x)$ given in Eq.~(\ref{DR}), the second term whose magnitude decreases rapidly when $z_R$ increases from $-\sqrt2 N_R$ towards $-\nu_R$ is dominant over the first one in the asymptotic region  for  moderate $\frac{1}{|\sin(\nu_R\pi)|}$, in accordance with the localization. 
This localization  in turn suggests that the corresponding eigenstate is non-degenerate.
 Indeed, localized non-degenerate eigenstates have been found numerically in various asymmetric double-well systems without necessarily assuming parabolicities of the wells \cite{WT,JPSADLW}. Likewise, for  the non-degenerate eigenstate of $E\approx \epsilon_R$ with moderate $\frac{1}{|\sin(\nu_L\pi)|}$,  Eq.~($\ref{LR1}$) shows that the eigenstate is primarily localized in the right-hand well.

If  $ \epsilon_L\approx \epsilon_R$ so that both Eqs.~(\ref{LR2}) and (\ref{LR1}) are valid, they give rise to the quantization condition  for the energy levels:
\begin{equation}
\tan(\pi\nu_L) \tan(\pi\nu_R)=\frac{1}{4}g_{\nu_L} g_{\nu_R} \exp\left(-2\int_{a_{\nu_L}}^{a_{\nu_R}}\frac{p(y)}{\hbar}dy\right).
\label{cc}
\end{equation}
We note that quantization formula analogous to Eq.~(\ref{cc}) has been found  in a semiclassical analysis for a general  double-well potential problem \cite{FF,Connor};  in these previous derivations, however, since the wells are not assumed to be  parabolic, the equation has been given  in terms of the action integrals on the left-hand side and without the prefactor $g_{\nu_L} g_{\nu_R}$ on the right-hand side \cite{FF,Connor}.  
From its derivation given here, we note that Eq.~(\ref{cc}) is valid {\em only when}   $\nu_L\simeq n_L$ {\em and} 
$\nu_R\simeq n_R$; further, in this equation,
$\frac{1}{|\sin(\nu_R\pi)|}$ and $\frac{1}{|\sin(\nu_R\pi)|}$ are implicitly assumed to be moderately large  so that one term is not exponentially dominant over the other in the overlap region in Eqs.~(\ref{DL}) and (\ref{DR}),  for the validity of Eqs.~($\ref{ACL}$) and (\ref{BCR}) (see also next section).

For further analyses of the quantization condition, with the constants $\delta_{n_L},~\delta_{n_R}$ satisfying
\begin{equation}
\epsilon_L +\hbar \omega_L \delta_{n_L} =\epsilon_R +\hbar \omega_R \delta_{n_R},
\end{equation}
we define 
\begin{equation}
\triangle \epsilon =\epsilon_L - \epsilon_R =\hbar(\delta_{n_R} \omega_R -\delta_{n_L} \omega_L),
\end{equation}
to rewrite  $\nu_L$, $\nu_R$ in Eq.~(\ref{defenergy}) as
\begin{equation}
\nu_L=n_L+\delta_{n_L}+\delta_L,~~~~\nu_R=n_R+\delta_{n_R}+\frac{\omega_L}{\omega_R}\delta_L,
\label{defnuLR}
\end{equation}
with a constant $\delta_L$ ($|\delta_L|\ll 1$).  Equation ($\ref{cc}$) is then approximated as
\begin{equation}
(\delta_L+\frac{\omega_R}{\omega_L}\delta_{n_R})(\delta_L +\delta_{n_L})
=\frac{\omega_R}{\omega_L}\left[\frac{\sqrt{{g_{n_L} g_{n_R}}}}{2\pi}
 \exp\left(-\int_{a_{n_L}}^{a_{n_R}}\frac{p(y)}{\hbar}dy\right) \right]^2 .
\label{appcc}
\end{equation}
Two real roots of the quadratic equation of  Eq.~($\ref{appcc}$) for $\delta_L$ give  two energy eigenvalues $E_\pm$ .

For the degenerate left and right states of  $ \epsilon_R= \epsilon_L$, using $\delta_{n_L}=\delta_{n_R}=0$, we obtain  the splitting  ${\it\Delta}$ $(=2|\delta_L|\hbar\omega_L)$ between the two energy eigenvalues: 
\begin{equation}
{\it\Delta}=\frac{\hbar}{\pi} \sqrt{g_{n_L}g_{n_R} \omega_L \omega_R}
 \exp\left(-\int_{a_{n_L}}^{a_{n_R}}\frac{p(y)}{\hbar}dy\right) 
\label{itDelta}
\end{equation}
which exactly agrees with the result given in Ref.~\cite{JPSADLW}.
Using Eq.~(\ref{defnuLR}), 
$C_L/C_R$ can be 
approximated from Eqs.~(\ref{LR2}) (or (\ref{LR1})) as
\begin{equation}
\frac{C_L}{C_R}= \pm (-1)^{n_L+n_R}  \sqrt{\frac{n_R!l_R}{n_L!l_L}}
\label{LRapp}
\end{equation}
for $E=E_{\mp}$. 
The following formula for an integer $n$ \cite{WW}
\[\int_{-\infty}^\infty \{D_n(z)\}^2 dz =\sqrt{2\pi} n!,\]
and  Eq.~(\ref{LRapp})
then imply that the left-right amplitude ratio  is approximately  unity (that is, $\int_{-\infty}^c|\psi(x)|^2dx \simeq \int_c^{\infty}|\psi(x)|^2dx$)
for $E=E_{\pm}$.

For $\triangle \epsilon\neq 0$,
with the two  roots $\delta_{L_+}$ and $\delta_{L_-}$ of Eq.~($\ref{appcc}$), we find the eigenvalues  
\begin{eqnarray}
E_\pm&=&\hbar\omega_L\left[n_L +\delta_{n_L}+\frac{1}{2}
+\frac{1}{2}\left\{(\delta_{L_+} +\delta_{L_-})\pm(\delta_{L_+} -\delta_{L_-})\right\} \right] +V(a_L)\nonumber\\
&=&\hbar\omega_L\left[ n_L +\frac{1}{2}\left(1+\delta_{n_L}- \frac{\omega_R}{\omega_L}\delta_{n_L}\right)   \right]+V(a_L)
\pm\frac{1}{2}\sqrt{(\triangle \epsilon)^2 + {\it\Delta}^2}\nonumber\\
&=&\frac{1}{2}(\epsilon_L + \epsilon_R )\pm\frac{1}{2}    \sqrt{(\triangle \epsilon)^2 + {\it\Delta}^2},
\label{E_pm}
\end{eqnarray}
to get the energy splitting
\begin{equation}
\triangle E=E_+ -E_-=\sqrt{(\triangle \epsilon)^2+ {\it\Delta}^2}.
\label{TLS}
\end{equation}
In view of the discussions on two-level systems (see, e.g., Ref.~\cite{CDL}), Eq.~(\ref{TLS}) suggests that, for $\triangle E\ll \hbar\omega_L$ and $\triangle E\ll \hbar\omega_R$, we can confine ourselves  to a two-dimensional subspace described by the eigenfunctions corresponding to  the eigenvalues  $E=E_{\pm}$ as a first approximation.

\section{Two-level approximation and energy splitting }
\label{sec:two}

For a two-level approximation, we hypothesize two approximate solutions  to the Schr\"{o}dinger equation  $\tilde\psi_L (x)$ and $\tilde\psi_R (x)$
which are primarily localized with energy $E$ in the left-hand and  right-hand wells, respectively.
For $\tilde\psi_L (x)$ with $E\simeq\epsilon_L$,
we naturally approximate  $\tilde\psi_L (x)$ by $\psi_{n_L}^{sho}(a_L;x)$ near $x=  a_L$,  which will be more accurate in the (Hermite-Gaussian) limit  of $E\rightarrow\epsilon_L$ ($\nu_L\rightarrow n_L$).
The pertinent WKB wave function in the forbidden region between the wells is then
\begin{equation}
\psi_{WKB}^L(x)= N_L\sqrt{\frac{\hbar}{p(x)}}e^{-\int_c^x \frac{p(y)}{\hbar} dy}
\end{equation}
the amplitude of which decreases as $x$ increases.
In the region of quadratic potential satisfying $\sqrt{2}(-a_L+x)\gg l_L\sqrt{2n_L+1}$, using Eq.~(\ref{W_L}),
we have 
\begin{eqnarray}
&&\psi_{WKB}^L(x)= N_L\sqrt{\frac{\hbar}{p(x)}}\exp \left(\int_{a_{n_L}}^c\frac{p(y)}{\hbar} dy -W_L(n_L)\right)\phantom{\hspace{3.3cm}}
\cr
&&\simeq N_L\frac{(4\pi)^{\frac{1}{4}}\sqrt{l_L}}{\sqrt{ n_L!g_{n_L}}}
\left(\frac{\sqrt{2}(x-a_L)}{l_L}\right)^{n_L}
\exp\left(-\frac{(-a_L+x)^2}{2l_L^2}+\int_{a_{n_L}}^c \frac{p(y)}{\hbar}dy\right).
\end{eqnarray}
Requiring that this asymptotic form of $\psi_{WKB}^L(x)$ matches onto that of $\psi_{n_L}^{sho}(a_L;x)$ in the forbidden region near the turning point,
we find
\begin{equation}
N_L=\sqrt{\frac{ g_{n_L}}{2\pi}}\frac{1}{l_L}\exp\left[- \int_{a_{n_L}}^c\frac{p(y)}{\hbar} dy\right],
\label{N_L}
\end{equation}
and thus \cite{Song}
\begin{equation}
\psi_{WKB}^L(x)= \sqrt{\frac{m\omega_L g_{n_L}}{2\pi p(x)}}~e^{-\int_{a_{n_L}}^x \frac{p(y)}{\hbar} dy}.
\end{equation}
Since we are considering $\tilde\psi_L (x)$  primarily localized in the left-hand well,  we will  take  $\tilde\psi_L (x)=0$ for $x>a_R- N_R l_R$ in further use.
For $\tilde\psi_R (x)$ with $E\simeq\epsilon_R$,  the pertinent WKB wave function  in the forbidden region between the wells is
\begin{equation}
\psi_{WKB}^R(x)= N_R\sqrt{\frac{\hbar}{p(x)}}e^{\int_c^x \frac{p(y)}{\hbar} dy}.
\end{equation}
Requiring that the asymptotic form of $\psi_{WKB}^R(x)$ matches onto that of  $\psi_{n_R}^{sho}(a_R;x)$ in  the region of quadratic potential satisfying $\sqrt{2}(a_R-x)\gg l_R\sqrt{2n_R+1}$, where $\psi_{n_R}^{sho}(a_R;x)$ is defined by the replacement of the subscript $L$ by $R$ in Eq.~(\ref{sho}), we have
\begin{equation}
N_R=(-1)^{n_R}\sqrt{\frac{ g_{n_R}}{2\pi}}\frac{1}{l_R}\exp\left[- \int_c^{a_{n_R}}
\frac{p(y)}{\hbar} dy\right].
\label{N_R}
\end{equation}
The factor $(-1)^{n_R}$ in Eq.~(\ref{N_R}) is due to  the negativity of  the  harmonic oscillator wave function $\psi_{n_R}^{sho}(a_R;x)$ of  odd $n_R$ in the limit of $x-a_R\ll -l_R$.

Assuming that $\epsilon_L$ and $\epsilon_R$ are close together and  very different from the energy eigenvalues of all other states of the system,
we may restrict our attention on the  two-dimensional subspace spanned by $\tilde\psi_L (x)$ and  $\tilde\psi_R (x)$ \cite{CDL}.
For simplicity we start with the case of  $\epsilon_L=\epsilon_R=E_0$  in which  all the diagonal elements of the $2\times 2$ Hamiltonian  matrix in the subspace are $E_0$, and  the off-diagonal elements $\int_{-\infty}^\infty \tilde\psi_L^* (x)H\tilde\psi_R (x)$ and $\int_{-\infty}^\infty \tilde\psi_R^* (x)H\tilde\psi_L (x)$   are equal to each other since we have taken  $\tilde\psi_L (x)$ and $\tilde\psi_R (x)$ to be real. 
It is easy to find that the eigenfunctions corresponding to the  energy eigenvalues $E_0\pm\frac{\tilde{\it\Delta}}{2}$ in the subspace are written as
\begin{equation}
\tilde\psi_\pm(x)= \frac{1}{\sqrt2}\left(\tilde\psi_L (x)\mp (-1)^{n_R} \tilde\psi_R (x) \right),
\end{equation}
where
\begin{equation}
\tilde{\it\Delta}= 2\times(-1)^{n_R+1}\int_{-\infty}^\infty \tilde\psi_L (x)H\tilde\psi_R (x).
\label{defDelta}
\end{equation}
To find an explicit expression of $\tilde{\it\Delta}$, as in Ref.~\cite{DekkerPRA}, we multiply the Schr\"{o}dinger equation ($\ref{Schroedinger}$) for $\tilde\psi_+(x)$ with energy $E_0+\frac{\tilde{\it\Delta}}{2}$ by  $\tilde\psi_-(x)$, and the  equation for $\tilde\psi_-(x)$ with energy $E_0-\frac{\tilde{\it\Delta}}{2}$ by $\tilde\psi_+(x)$. Subtracting the two resulting expressions and integrating over $x\in [c,\infty)$, with the facts 
\begin{equation}
\int_c^\infty (\tilde\psi_R (x))^2 dx\simeq 1,~~\int_{c}^\infty (\tilde\psi_L (x))^2 dx\simeq 0,~~{\rm and}~\int_{c}^\infty \tilde\psi_L (x) \tilde\psi_R (x) dx \simeq 0
\end{equation}
which are given by the definitions of $\tilde\psi_L (x)$ and $\tilde\psi_R (x)$,
we find 
\begin{equation}
\tilde{\it\Delta}\simeq (-1)^{n_R}\frac{2\hbar^2}{m}N_LN_R.
\label{twoDelta}
\end{equation}
With the explicit expressions given in Eqs.~(\ref{N_L}) and (\ref{N_R}),  it is easy to check that $\tilde{\it\Delta}$ reduces to ${\it\Delta}$ of Eq.~(\ref{itDelta}).
Moreover, using the expressions of $A$ and $B$ in Eqs.~(\ref{ACL}) and (\ref{BCL}) (or, in  Eqs.~(\ref{ACR}) and (\ref{BCR})), if  $\nu_L=n_L \pm \frac{\it\Delta}{2\hbar \omega_L}$ and $\nu_R=n_R \pm \frac{\it\Delta}{2\hbar \omega_R}$,  we find  that
\begin{equation}
\frac{B}{A}\simeq\mp(-1)^{n_R} \frac{N_L}{N_R}.
\label{ABratio}
\end{equation}
Equation (\ref{ABratio}) implies that  $\tilde\psi_\pm(x)$ reduces  to   $\psi(x)$ for $E=E_\pm$ up to  normalization constants in the forbidden region between the wells, to show that this two-level description is indeed equivalent to the wave function approach of the previous section, for $\triangle \epsilon =0$.

In the (Hermite-Gaussian) limit of $\nu_L\rightarrow n_L$ Eq.~(\ref{N_L}) is equal to Eq.~(\ref{BCL}) with $B=N_L$ and $C_L=(-1)^{n_L}\frac{1}{\pi^{\frac{1}{4}}\sqrt{n_L! l_L}}$, and  in the limit of $\nu_R\rightarrow n_R$  Eq.~(\ref{N_R}) is equal to Eq.~(\ref{ACR}) when $A=N_R$ and $C_R=\frac{1}{\pi^{\frac{1}{4}}\sqrt{n_R! l_R}}$. 
For the low-lying states, the approximation of $\tilde\psi_L (x)$  ($\tilde\psi_R (x)$) by  $\psi_{n_L}^{sho}(a_L;x)$  ($\psi_{n_R}^{sho}(a_R;x)) $ near the turning point would be more accurate in the limit of $\nu_L\rightarrow n_L$ ($\nu_R\rightarrow n_R$), while, as has been discussed  in the previous section, the WKB method does not provide the validity of Eq.~(\ref{ACL}) (Eq.~(\ref{BCR})) in the limit;
indeed,  neither Eq.~(\ref{ACL}) nor  Eq.~(\ref{BCR}) has appeared in obtaining Eq.~(\ref{twoDelta}). 
Nevertheless,  the fact $\tilde{\it\Delta}\simeq{\it\Delta}$ may suggest that  when  $\nu_L\simeq n_L$ and 
$\nu_R\simeq n_R$, $\nu_L$  and $\nu_R$  adjust themselves so that  $\frac{1}{|\sin(\nu_L\pi)|}$  and $\frac{1}{|\sin(\nu_R\pi)|}$ are moderately large and  thus Eqs.~(\ref{ACL}) and (\ref{BCR}) are in fact valid to render Dekker's method applicable.

It is straightforward to extend the description to include the case of  $\triangle \epsilon \neq 0$. Without losing generality, we assume $\triangle \epsilon  > 0$. In this case, the $2\times 2$ Hamiltonian matrix  in the subspace spanned by $\tilde\psi_L (x)$ and  $\tilde\psi_R (x)$ is 
\begin{eqnarray}
&&\left(\begin{array}{cc}
\epsilon_L  &  (-1)^{n_R+1}\frac{\tilde{\it\Delta}}{2} \\
 (-1)^{n_R+1}\frac{\tilde{\it\Delta}}{2}    &  \epsilon_R           \end{array}\right)\phantom{\hspace{6cm}} \cr
&&~~=
\frac{\epsilon_L+\epsilon_R}{2}
\left(\begin{array}{cc}
1  &  0 \\
0  &  1           \end{array}\right)
+\frac{1}{2}\sqrt{(\triangle \epsilon)^2+ {\tilde{\it\Delta}}^2}
\left(\begin{array}{cc}
\cos \theta  &   \sin \theta  \\
\sin \theta  &  -\cos \theta           \end{array}\right),
\end{eqnarray}
and the eigenfunctions $\psi_\pm(\theta;x)$ of the Hamiltonian corresponding to the eigenvalues $E_\pm$ of Eq.~(\ref{E_pm}) are written as
\begin{eqnarray}
\psi_+(\theta;x)&=&\cos (\frac{\theta}{2}) \tilde\psi_L(x) +   \sin (\frac{\theta}{2}) \tilde\psi_R(x),\cr
\psi_-(\theta;x)&=&(-1)^{n_R}\left[-\sin (\frac{\theta}{2}) \tilde\psi_L(x) +   \cos (\frac{\theta}{2})\tilde \psi_R(x)\right]
\end{eqnarray}
up to constant phases, where 
\begin{equation}
\cos \theta = \frac{\triangle \epsilon}{\sqrt{(\triangle \epsilon)^2+ {\tilde{\it\Delta}}^2}},~~
\sin \theta = \frac{(-1)^{n_R+1}\tilde{\it\Delta}}{\sqrt{(\triangle \epsilon)^2+ {\tilde{\it\Delta}}^2}}
\end{equation}
with $-\frac{\pi}{2}< \theta < \frac{\pi}{2}$. 
The method described in Ref.~\cite{DekkerPRA} can also be used as follows;
\begin{eqnarray}
&&\int_c^\infty[\psi_-(\theta;x)H\psi_+(\theta;x)-\psi_+(\theta;x)H\psi_-(\theta;x)]dx\cr
&&=\sqrt{\triangle \epsilon)^2+ {\tilde{\it\Delta}}^2}\int_c^\infty \psi_-(\theta;x)\psi_+(\theta;x)dx
\simeq -\frac{\tilde{\it\Delta}}{2}\\
&&=-\frac{\hbar^2}{2m}\int_c^\infty\frac{d}{dx}\left[\psi_-(\theta;x)\frac{d}{dx}\psi_+(\theta;x)-\psi_+(\theta;x)\frac{d}{dx}\psi_-(\theta;x)\right]dx\cr
&&=(-1)^{n_R}~\frac{\hbar^2}{2m}\int_c^\infty\frac{d}{dx}\left[\psi_L(x)\frac{d}{dx}\psi_R(x)-\psi_R(x)\frac{d}{dx}\psi_L(x)\right]dx\cr
&&=(-1)^{n_R+1}~\frac{\hbar^2}{2m}\left[\psi_L(x)\frac{d}{dx}\psi_R(x)-\psi_R(x)\frac{d}{dx}\psi_L(x)\right]\bigg|_{x=c}\cr
&&=(-1)^{n_R+1}\frac{\hbar^2}{m}N_LN_R,
\label{DekkerM}
\end{eqnarray}
to show that the expression of $\tilde{\it\Delta}$ in Eq.~(\ref{twoDelta}) is still valid in this case. In obtaining the last equality in Eq.~(\ref{DekkerM}), we used the fact that $V(x)$ is differentiable at $x=c$.
It is easy to check that,  if we take the limit of $\theta\rightarrow (-1)^{n_R+1}\frac{\pi}{2}$, $\psi_\pm(\theta;x)$ reduce to $\tilde\psi_\pm(x)$ of $\triangle \epsilon =0$.

\section{ Concluding remarks}
\label{sec:concluding}
We have systematically applied Dekker's method for the low-lying states of an asymmetric potential. While the method gives the quantization condition for  two almost degenerate states which reproduces the energy splitting formula obtained through the two-level approach \cite{JPSADLW}, for the other low-lying states it shows that the eigenstate is non-degenerate and  primarily localized in one of the wells.
Indeed, numerical studies with \cite{WT} or without parabolicity  \cite{JPSADLW} implies that the low-lying eigenstate of an asymmetric double-well potential is either one of the two almost degenerate tunneling states or a localized non-degenerate state, in accordance with the analytic reasons given  here. 
Though   we have introduced $c$ satisfying $a_L+N_Ll_L < c< a_R-N_R l_R$ during the derivations, the point $c$  appears neither in  the quantization condition nor in $C_L/C_R$. This independence from the choice of $c$ implies that the splitting formula in Eqs.~(\ref{itDelta}) and (\ref{TLS}) would be applicable even when the potential barrier is not simply concave downwards. It would be interesting to test the splitting formula for the potential which is above $V(a_{n_L})$ $(=V(a_{n_R}))$ between $a_{n_L}$ and $a_{n_R}$ but has structures such as bumps and wells. Though numerical methods can be  used  for a specific one-dimensional potential problem to find the eigenvalues, however, we believe that the analytic result which is shown to be accurate  for the low-lying states  in the numerical calculations \cite{JPSADLW} would be of importance by itself. 
The evaluation of the energy splitting in a symmetric double-well potential is known to be closely related to that of widths of energy bands in a periodic potential \cite{Coleman}, and  it would be interesting if one could extend Dekker's method to be applicable for a periodic problem.






\end{document}